# A quantum mechanical NMR simulation algorithm for protein-scale spin systems

Luke J. Edwards[a,b], D.V. Savostyanov[b],
Z.T. Welderufael[b], Donghan Lee[c], Ilya Kuprov[b,*]

[a]*Physical and Theoretical Chemistry Laboratory, Department of Chemistry, University of Oxford, South Parks Road, Oxford, OX1 3QZ, UK.*

[b]*School of Chemistry, University of Southampton, Highfield Campus, Southampton, SO17 1BJ, UK.*

[c]*Department of NMR based Structural Biology, Max Planck Institute for Biophysical Chemistry, Am Fassberg 11, D-37077 Goettingen, Germany.*

Fax: +44 2380 594140

Email: i.kuprov@soton.ac.uk

**Abstract**

Nuclear magnetic resonance spectroscopy is one of the few remaining areas of physical chemistry for which polynomially scaling simulation methods have not so far been available. Here, we report such a method and illustrate its performance by simulating common 2D and 3D liquid state NMR experiments (including accurate description of spin relaxation processes) on isotopically enriched human ubiquitin – a protein containing over a thousand nuclear spins forming an irregular polycyclic three-dimensional coupling lattice. The algorithm uses careful tailoring of the density operator space to only include nuclear spin states that are populated to a significant extent. The reduced state space is generated by analysing spin connectivity and decoherence properties: rapidly relaxing states as well as correlations between topologically remote spins are dropped from the basis set. In the examples provided, the resulting reduction in the quantum mechanical simulation time is by many orders of magnitude.

**Introduction**

The computing power required for nuclear magnetic resonance (NMR) simulations grows exponentially with the spin system size[1], and the current simulation capability is limited to about twenty spins[2]. Proteins are much bigger and the inability to accurately model their NMR spectra is a significant limitation. In particular, exponential scaling complicates validation of protein NMR structures: an *ab initio* simulation of a protein NMR spectrum from atomic coordinates and list of spin interactions has not so far been feasible. It is also not possible to cut a protein up into fragments and simulate it piecewise without losing essential dipolar network information[3]. For this reason, some of the most informative protein NMR experiments (*e.g.* NOESY) are currently only interpreted using simplified models[4]. Very promising recent algorithms, such as DMRG[5], are also challenged by time-domain NMR simulations of proteins, which contain irregular three-dimensional polycyclic spin-spin coupling networks that are far from chain or tree topologies required by tensor network methods. In this communication we take advantage of the locality and rapid relaxation properties of protein spin systems and report a solution to the protein NMR simulation problem using restricted state spaces[6]. NOESY, HNCO and HSQC simulations of $^{13}C,^{15}N$-enriched human ubiquitin protein (over 1000 coupled spins) are provided as illustrations.

**Simulation methods**

The restricted state space approximation in magnetic resonance[6] is the observation that a large part of the density operator space in many spin systems remains unpopulated and can be ignored – the analysis of quantum trajectories in liquid state NMR indicates that only low orders of correlation connecting nearby spins are in practice populated[6,7]. The reasons, recently explored[6-14], include sparsity of common spin interaction networks[6,7], the inevitable presence of spin relaxation[11,15], the existence of multiple non-interacting density matrix subspaces[10,12], the presence of hidden conservation laws[12] and simplifications brought about by the powder averaging operation[8,14]. It is possible to determine the composition of the reduced space *a priori*, allowing the matrix representations of spin operators to be built directly in the reduced basis set[11,12]. Taken together, this yields a polynomially scaling method for simulating liquid phase NMR systems of arbitrary size. Our final version of this method is described in this communication – we build the reduced operator algebra by only including populated spin product states in the basis. The populated subspace is mapped by analysing the topology of the spin interaction network. A rigorous accuracy analysis is highly technical and has been published separately[11].

There are two distinct spin interaction networks in NMR systems: the *J*-coupling network, defined by electron-mediated interactions that propagate through chemical bonds, and the dipolar coupling network, defined by through-space magnetic dipolar couplings between nuclei. In the liquid phase, these two networks have very different manifestations: the *J*-coupling network is responsible for multiplicity patterns observed directly in NMR spectra, whereas the dipolar network is partially responsible for line widths and cross-relaxation processes. Both networks are irregular, three-dimensional, and contain multiple interlocking loops that challenge current DMRG techniques[5]. In a typical NMR experiment, nuclear magnetization flows across both networks and the locality of the operator basis set should therefore be understood as locality on the corresponding graphs.

After testing a variety of state space restriction methods[6,7,11-14], we propose the following procedure for generating the reduced basis set in liquid state NMR simulations:

1. Generate *J*-coupling graph (JCG) and dipolar coupling graph (DCG) from *J*-coupling data and Cartesian coordinates respectively. User-specified thresholds should be applied for the minimum significant *J*-coupling and maximum significant distance. Because spin interactions are at most two-particle, the computational complexity of this procedure and the number of edges in the resulting graphs scale quadratically with the number of spins.

2. Use the depth-first search algorithm[16] on both JCG and DCG to generate the complete list of connected overlapping subgraphs involving a user-specified number of spins. This number controls the approximation accuracy[11] and should be specified independently for JCG and DCG. The complexity of this procedure and the number of the resulting subgraphs scale linearly with the number of edges in JCG and DCG[16].

3. For each subgraph $G_k$, generate a description of the complete basis set of the corresponding spin subsystem. The dimension $D_k$ of this basis set is equal to the product of squares of multiplicities of each spin in $G_k$ and does not depend on the size of the overall spin system. The most convenient operator basis set is direct products of irreducible spherical tensors, where each basis operator has the following structure:

$$\hat{T}_{l_1,m_1} \otimes \hat{T}_{l_2,m_2} \otimes \ldots \otimes \hat{T}_{l_j,m_j} \otimes \ldots \otimes \hat{T}_{l_{|G_k|},m_{|G_k|}} \qquad (1)$$

where $\hat{T}_{l_j,m_j}$ is an irreducible spherical tensor operator of rank $l_j$ and projection $m_j$ acting on spin $j$ and $|G_k|$ is the number of spins in subgraph $G_k$. A useful feature of Equation (1) is that matrix storage is avoided because the structure of each basis operator is completely determined by the index sequence $\{l_j, m_j\}$. Therefore, the description of the complete operator basis of a given subgraph $G_k$ requires $2|G_k|D_k$ integers of storage space and does not depend on the total number of spins in the system. The corresponding list of basis operator descriptors will henceforth be referred to as the "state list".

4. Merge state lists of all subgraphs and eliminate repetitions caused by subgraph overlap. This procedure results in a basis set that contains only low orders of spin correlation (by construction, up to the size of the biggest subgraph) between spins that are proximate on JCG and DCG (by construction, because connected subgraphs were generated in Stage 2). At the same time, the resulting basis describes the entire system without gaps or cuts: once the subgraph state lists are merged and repetitions are eliminated, the result is a global list of spin operators that are expected to be populated during the spin system evolution based on the proposed heuristics of locality and low correlation order.

The accuracy of the basis set can be varied systematically by changing subgraph size in Stage 2 – the limiting case of the whole system corresponds to the formally exact simulation[11]. The basis set nomenclature implemented in our software library, called *Spinach*[17], and used for the simulations described below, is given in Table 1. The procedure described above runs in quadratic time with respect to the total number of spins in the system.

Once the active space is mapped, matrix representations should be built for relevant spin operators and state vectors. Experimentally encountered spin interactions are at most two-particle, and the Hamiltonian appearing in the equation of motion for the density operator is therefore a sum of at most two-spin operators with a known direct product structure[1]:

$$\hat{H}_n = \omega_n \left[ \hat{\sigma}_{n,1} \otimes \hat{\sigma}_{n,2} \otimes \cdots \otimes \hat{\sigma}_{n,N} \right] \qquad (2)$$

where $\omega_n$ are interaction magnitudes, $N$ is the total number of spins, and $\hat{\sigma}_{n,k}$ are identity matrices, Pauli matrices or spherical tensor operators of dimension $2s_k + 1$ in which $s_k$ is the spin quantum number of *k*-th nucleus. The corresponding commutation superoperators $\hat{\hat{H}}_n^{(C)}$ can be written as differences between left-side and right-side product superoperators $\hat{\hat{H}}_n^{(L)}$ and $\hat{\hat{H}}_n^{(R)}$, defined by their action on a density operator $\hat{\rho}$:

$$\hat{\hat{H}}^{(C)} = \sum_n \hat{\hat{H}}_n^{(C)} = \sum_n \left( \hat{\hat{H}}_n^{(L)} - \hat{\hat{H}}_n^{(R)} \right)$$

$$\hat{\hat{H}}_n^{(C)} \hat{\rho} = \left[ \hat{H}_n, \hat{\rho} \right] = \hat{H}_n \hat{\rho} - \hat{\rho} \hat{H}_n \qquad \hat{\hat{H}}_n^{(L)} \hat{\rho} = \hat{H}_n \hat{\rho} \qquad \hat{\hat{H}}_n^{(R)} \hat{\rho} = \hat{\rho} \hat{H}_n \qquad (3)$$

Their faithful representations have exponential dimensions, but representations in low correlation order basis sets are cheap[12]. In a given operator basis $\{\hat{O}_k\}$:

$$\left[ \hat{\hat{H}}_n^{(L)} \right]_{jk} = \left\langle \hat{O}_j \middle| \hat{\hat{H}}_n^{(L)} \middle| \hat{O}_k \right\rangle = \mathrm{Tr}\left[ \hat{O}_j^\dagger \hat{H}_n \hat{O}_k \right] = \mathrm{Tr}\left[ \left( \bigotimes_{m=1}^N \hat{\sigma}_{j,m}^\dagger \right) \left( \bigotimes_{m=1}^N \hat{\sigma}_{n,m} \right) \left( \bigotimes_{m=1}^N \hat{\sigma}_{k,m} \right) \right] \qquad (4)$$

Because dot products commute with direct products and the trace of a direct product is a product of traces we have:

$$\left[ \hat{\hat{H}}_n^{(L)} \right]_{jk} = \mathrm{Tr}\left[ \bigotimes_{m=1}^N \left( \hat{\sigma}_{j,m}^\dagger \hat{\sigma}_{n,m} \hat{\sigma}_{k,m} \right) \right] = \prod_{m=1}^N \mathrm{Tr}\left[ \hat{\sigma}_{j,m}^\dagger \hat{\sigma}_{n,m} \hat{\sigma}_{k,m} \right] \qquad (5)$$

in which the dimension of individual matrices $\hat{\sigma}_{n,k}$ is tiny and does not depend on the size of the spin system; the computational complexity of computing $\mathrm{Tr}\left[ \hat{\sigma}_{j,m}^\dagger \hat{\sigma}_{n,m} \hat{\sigma}_{k,m} \right]$ is therefore $O(1)$ and the complexity of computing one matrix element is $O(N)$ multiplications, where $N$ is the total number of spins in the system. With $O(N^2)$ interactions in the spin system, this puts the worst-case complexity of building the representation of the Hamiltonian in Equation (3) to $O(N^3 D^2)$, where $D$ is the dimension of the reduced basis set. The sparsity of spin Hamiltonians[18] and the fact that spin interaction networks in proteins are also sparse puts the practically observed scaling closer to $O(N^2 D)$ – a significant improvement on the $O(4^N)$ best-case scaling of the adjoint direct product representation. This improvement is further amplified by the presence of unpopulated states even in the low correlation order subspace[7], by the existence of multiple independently evolving subspaces[12], and by the fact that not all of the populated states belong to the propagator group orbit of the detection state[10].

Matrix dimension, storage and CPU time statistics for a 512×512 point $^1$H-$^1$H NOESY simulation of ubiquitin (573 protons, ~50,000 terms in the dipolar Hamiltonian) are given in Table 2. As demonstrated in Figures 1 and 2, the simulation is in good agreement with the experimental data. The state space restriction approximation reduces the Hamiltonian superoperator dimension from $4^{573} \approx 10^{345}$ to 848,530. The reduced Hamiltonian is still sparse, and therefore within reach of modern matrix manipulation techniques – the simulation shown in Figure 1 took less than 24 hours on a large shared-memory computer. Importantly, the problem dimension remains too big for matrix factorizations: the recently developed diagonalization-free methods[15] are essential.

The storage of the system trajectory in the indirect dimension of the 2D NMR simulation shown in Figure 1 requires 512×848,530 complex doubles (6.96 GB) of memory. It is clear that 3D NMR simulations would put some strain on modern computing facilities. This would have been a difficult problem, were it not for a peculiar property of propagator semigroups – simulations can be partially run backwards, even in the presence of relaxation. The general algebraic summary is given below and a special case of the HNCO pulse sequence is illustrated in Figure 3.

The free induction decay coming out of a 3D NMR experiment is a function of the three evolution times $\{t_1, t_2, t_3\}$ and may be formally written as

$$f(t_1,t_2,t_3) = \langle \hat{\sigma} | e^{-i\hat{L}t_3} \hat{P}_3 e^{-\frac{i\hat{L}t_2}{2}} \hat{M}_2 e^{-\frac{i\hat{L}t_2}{2}} \hat{P}_2 e^{-\frac{i\hat{L}t_1}{2}} \hat{M}_1 e^{-\frac{i\hat{L}t_1}{2}} \hat{P}_1 | \hat{\rho}_0 \rangle, \qquad \hat{L} = \hat{H} + i\hat{R} \qquad (6)$$

where $|\hat{\rho}_0\rangle$ is the initial density matrix, $|\hat{\sigma}\rangle$ is the detection state, $\hat{L}$ is the background Liouvillian of the system comprising a Hamiltonian $\hat{H}$ and a relaxation superoperator $\hat{R}$, $\hat{P}_n$ are preparation pulse and delay propagators, and $\hat{M}_n$ are propagators of refocusing pulses in the middle of evolution periods. Because semigroups are associative, the result of Equation (6) does not depend on the partitioning of Dirac brackets. In particular,

$$f(t_1,t_2,t_3) = \left\langle \hat{\sigma} e^{-i\hat{L}t_3} \hat{P}_3 e^{-\frac{i\hat{L}t_2}{2}} \hat{M}_2 \middle| e^{-\frac{i\hat{L}t_2}{2}} \hat{P}_2 e^{-\frac{i\hat{L}t_1}{2}} \hat{M}_1 e^{-\frac{i\hat{L}t_1}{2}} \hat{P}_1 \hat{\rho}_0 \right\rangle \qquad (7)$$

This transformation splits a 3D NMR simulation into one forward 2D simulation from the initial state, one backward 2D simulation from the detection state and one dot product in the middle. Equation (7) is formally equivalent to Equation (6), but the reduction in storage requirements is considerable – for a typical protein 3D NMR experiment, instead of a dense 64×64×256×10^6 array of complex doubles (over 16 TB of data) at the end of the $t_3$ period in Equation (6), the arrays in Equation (7) have dimensions of 64×64×10^6 and 64×256×10^6 as well as better sparsity, resulting in the worst-case storage requirements of about 256 GB. As per Equation (7), their scalar product along the last dimension returns the required 64×64×256 free induction decay. Importantly, Equation (7) retains the parallelization opportunities and the time-memory trade-offs offered by the fact that different $t_1$ increments may be evolved independently in $t_2$ forward, and different $t_3$ increments may be evolved independently in $t_2$ backward. The final operation – the matrix dot product in Equation (7) – is also intrinsically parallel. Practical testing shows that the two-sided propagation technique reduces the simulation time of 3D NMR experiments on proteins (HNCO example is given in Figure 4) by at least an order of magnitude.

Even in reduced spaces the algebraic structure of the time-domain NMR simulation problem lends itself to multiple efficiency tweaks. Sparse matrix algebra[19] is advantageous because in the Pauli basis all spin Hamiltonian matrices are guaranteed to be sparse[18]. The direct product structure in Equation (2) is completely defined by its indices – repeated requests for the same operator can be served from disk or RAM using the index array as a database record identifier. Parallelization is straightforward at both the propagation[18] and the housekeeping stages – individual operators in the Hamiltonian can be generated independently, there are 625 independent integrals in the relaxation superoperator[15] and hundreds of independently evolving subspaces during spin system evolution[12]. Another order of magnitude in simulation time is saved by replacing phase cycles with analytical coherence order selection – when the spherical tensor basis set is used, orders of spin coherence are the quantum numbers used to classify basis vectors, meaning that coherence order filters amount to zeroing the coefficients of the unwanted states. This removes the need to emulate spectrometer phase cycles, saving a factor of 8, 16 or 32 (depending on the phase cycle length) in the simulation time. After all of these refinements are applied, ubiquitin simulations run in about 24 hours.

**Experimental methods and data processing**

All NMR spectra were recorded at 300 K on Bruker AVANCE-III 900 and Varian Inova600 spectrometers equipped with $^1$H,$^{13}$C,$^{15}$N triple-resonance probes. 8.0 mM solution of $^{13}$C,$^{15}$N labelled human ubiquitin in D$_2$O, buffered at pH=5.8 (uncorrected for deuterium isotope effect) with 50 mM phosphate buffer, was used in all experiments. All related compounds were obtained commercially and used without further purification. NOESY[20], HNCO[21] and HSQC[22] spectra were recorded as described in the papers cited. NMR signal acquisition and digital signal processing parameters (window functions, time-domain zerofilling, frequency offsets) between the theoretical simulations and the experimental data were matched. Simulation source code listing the specific parameter values used is available at http://spindynamics.org as a part of the *Spinach* package[17] example set.

Currently available database records of protein chemical shifts are not complete[23,24] – rapidly exchanging protons, quaternary carbons and side chain nitrogens are often missing. The gaps in the chemical shift information were filled using literature average values reported by the BMRB database[24]. The following chemical shift data post-processing was then applied: symmetry-related methyl group protons (listed once in BMRB) were replicated using PDB coordinates; unassigned capping groups on C- and N-termini were ignored; all oxygen and sulphur atoms were removed ($^{16}$O, $^{32}$S and $^{34}$S nuclei have no spin); symmetry-related carbons and protons in PHE and TYR aromatic rings (listed once in BMRB) were replicated using PDB coordinates; protons of deuterated or exchanging groups, such as –OH or –NH$_3^+$, were ignored; magnetically equivalent –CH$_2$– group protons (listed once in BMRB) were replicated using PDB coordinates. The amplitude and orientation of the anisotropic parts of chemical shift tensors were assigned to backbone nitrogen and carbon spins (for which the correct description of CSA is essential) from literature data[25-27]. Spin relaxation in the amino acid side chains was assumed to be dipole-dipole dominated. *Matlab* code listing the specific parameter values used is available as a part of the *Spinach* package[17].

While chemical shift data is a necessary outcome of NMR structure determination[3], complete *J*-coupling data is not expected to be available in the foreseeable future for any protein. We found that missing *J*-couplings can be obtained with sufficient accuracy (±25% is required for 2D/3D NMR simulations reported) from atomic coordinates using semi-empirical estimates, and implemented a graph-theoretical estimator with the following stages:

1. The molecular bonding graph is partitioned into connected subgraphs of size two, and one-bond *J*-couplings are assigned from a complete database of atom pairs. Our experience with ubiquitin indicates that there are fewer than 100 unique connected atom pairs in regular proteins, and that most one-bond *J*-couplings within those pairs can be either found in the literature[3], or measured in individual amino acids, or estimated with sufficient accuracy using electronic structure theory software[28].

2. The molecular bonding graph is partitioned into connected subgraphs of size three, and two-bond *J*-couplings assigned from a complete database of connected atom triples. The number of unique connected atom triples in proteins is also reasonable – we saw fewer than 150 in regular proteins, a small enough number for an exhaustive list to be compiled from experiments, literature and electronic structure theory estimates.

3. The molecular bonding graph is partitioned into sequentially connected subgraphs of size four and dihedral angles are computed from atomic coordinates, allowing three-bond *J*-couplings to be assigned from a complete database of Karplus curves[29], with angle averaging for sites designated as mobile. Karplus curves are a well-researched topic, with specific data available for the backbone and less accurate generic curves available for rest of the structure[3]. The number of unique sequentially connected atom quartets found in proteins (fewer than 300, many belonging to similar structural types) was sufficiently small for a complete database of Karplus curves to be compiled from literature data, experiments, and electronic structure theory estimates.

*J*-couplings across more than three bonds were ignored. The effect of the electrostatic environment was also ignored – on the scale of the accuracy required for protein simulations its effect on *J*-coupling is small[30,31]. *Matlab* code listing the specific parameter values is available as a part of the *Spinach* package[17]. More accurate *J*-coupling estimation methods are undoubtedly possible, but are beyond the scope of the present paper – we should note very clearly here that *this paper is an exercise in quantum mechanics rather than structural biology*.

**Results and discussion**

Figures 1, 2, 4 and 5 illustrate the quantitative agreement of the simulation results with experimental data. The few missing peaks in Figures 4 and 5 correspond to either atoms missing from the database record or to spectral folding artefacts in the experimental data. The extra peaks appearing in the theoretical spectra correspond to protons that are rapidly exchanging with the deuterium of the solvent and therefore invisible in proton NMR experiment. The good agreement of the major NOESY cross-peak positions is apparent in Figure 1. The observed residual scatter in NOESY cross-peak intensities shown in Figure 2 is due to the following factors, whose detailed investigation we are leaving for future research:

1. A single set of atomic coordinates being used for the simulation. NMR structure

determination runs produce structural ensembles with dozens or hundreds of molecular geometries consistent with a given NMR data set. Running protein-scale NMR simulations on a molecular dynamics ensemble would require much greater computational resources, but is likely to reduce the point scatter observed in Figure 2.

2. A single global rotational diffusion tensor being used in the relaxation theory model in the simulations that produced Figures 1, 2, 4 and 5 – a stochastic Liouville equation add-on[32] would likely produce a better fit. Lipari-Szabo restricted local motion models[4] are another possibility – for the purposes of *ab initio* protein NMR simulations, the relevant local motion parameters may be extracted from molecular dynamics data[33].

Simulations shown in Figures 1-5 are currently on the brink of impossibility (over 500 GB of RAM is required), but the results are encouraging – liquid state NMR spectra of realistic protein spin systems can now be simulated. The following research avenues are now open:

1. Whole-protein optimization and benchmarking of NMR pulse sequences. We have published our preliminary research on the subject, dealing with a small fragment[34] – the algorithms described above enable protein-scale effort in that direction.

2. Optimal control optimization of biomolecular NMR experiments. The software library implementing protein-scale simulation algorithms already includes an Optimal control module[35,36] – it is now possible to adapt it for HSQC, HNCO, HNCOCA and other protein NMR pulse sequences.

3. Automatic protein NMR structure validation. Structure validation can be defined as making sure that atomic coordinates coming out of a crystallographic or NMR experiment correspond to reality and eliminating any mismatches between the mathematical solution and the true biological structure. The critical step in that process – back-calculation of protein NMR spectra – is now possible.

Taking a more distant and speculative view, it may eventually become feasible to run protein NMR structure determination and validation directly from atomic coordinates, using *ab initio* or DFT methods to predict spin interaction parameters and then the methods described above to generate candidate NMR spectra for least squares fitting. Such "direct structure fitting" has been demonstrated for EPR of small molecules[37]. Its routine use would require significant im-provements in the accuracy of quantum chemistry methods, but such improvements are quite likely in the next ten years.

**Conclusions**

The algorithm reported results in the reduction of liquid state NMR simulation time of protein-scale spin systems by many orders of magnitude – a considerable improvement over brute-force simulations using traditional techniques[1,19]. The method reported above does not require the spin system to be linear or regular, and does not require any modifications to the existing simulation code – the reduced operator matrices are drop-in replacements of their full-dimensional counterparts in the direct product formalism[1]. All procedures and examples described above are available as a part of our *Spinach* software library[17].

**Acknowledgements**

The project is supported by EPSRC (EP/F065205/1, EP/H003789/1, EP/J013080/1). The authors are grateful to Garnet K.-L. Chan, Christian Griesinger, Robert Laverick and Arthur G. Palmer for stimulating discussions.

# Figure captions

**Figure 1.** Experimental (left panels) and theoretical (right panels) $^1$H-$^1$H NOESY spectrum of ubiquitin at 900 MHz proton frequency with a mixing time of 65 ms. The simulated spectrum was obtained with the distance cut-off for dipolar interactions set to 4.0 Ångstrom. The relaxation superoperator (Bloch-Redfield-Wangsness theory with a single global rotational correlation time of 5 ns) was obtained with a diagonalization-free direct integration algorithm[15].

**Figure 2.** Correlation between experimental and theoretical $^1$H-$^1$H NOESY cross-peak volumes for ubiquitin at 900 MHz proton frequency. The relaxation superoperator (Bloch-Redfield-Wangsness theory with a single global rotational correlation time of 5 ns) was obtained with a diagonalization-free direct integration algorithm[15].

**Figure 3.** Bidirectional propagation method schematic for the simulation of 3D HNCO NMR experiment[21]. Time is run forward from the initial condition to the middle of the $t_2$ period and backward from the detection state to the middle of the $t_2$ period. Both halves have the computational complexity of a 2D simulation and their scalar product generates the required 3D free induction decay. The channel labelled M represents analytical coherence selection and decoupling that are achieved by directly modifying the system state vector or Hamiltonian.

**Figure 4.** Theoretical (right panel) and experimental (left panel) 3D HNCO NMR spectra of $^{13}$C, $^{15}$N labelled ubiquitin, obtained using the pulse sequence shown in Fig. S1. The minor differences in peak intensities are due to non-uniform partial deuteration of the protein by the solvent as well as: (**A**) aliasing of arginine $N_\varepsilon$-$H_\varepsilon$ signals from their position at ~90 ppm $^{15}$N chemical shift which is outside the spectral window – simulated spectra are free of this artefact; (**B, C**) rapid exchange of $H_N$ protons in GLU24 and GLY53 with the deuterium of the solvent – the corresponding signals are lost in the noise in the experimental data.

**Figure 5.** Theoretical (left panel) and experimental (right panel) $^1$H-$^{15}$N HSQC NMR spectra of $^{13}$C, $^{15}$N labelled ubiquitin. The differences between the two spectra are due to: (**A, B**) rapid exchange of $H_N$ protons in GLU24 and GLY53 with the deuterium of the solvent – the corresponding signals are lost in the noise in the experimental data; (**C**) aliasing of arginine $N_\varepsilon$-$H_\varepsilon$ signals from their position at ~90 ppm $^{15}$N chemical shift that is outside the spectral window – simulated spectra are free of this artefact; (**D**) slow exchange of $H_\varepsilon$ protons in GLN41 with the deuterium of the solvent – the corresponding pair of signals is attenuated in the experiment.

**Table 1**: Reduced basis set nomenclature for liquid state NMR simulations implemented in *Spinach* library.

| Basis set | Description |
|---|---|
| IK-0($n$) | All spin correlations up to, and including, order $n$, irrespective of proximity on *J*-coupling or dipolar coupling graphs. Generated with a combinatorial procedure, by picking all possible groups of $n$ spins in the current spin system and merging state spaces of those groups. Recommended for testing and debugging purposes. |
| IK-1($n,k$) | All spin correlations up to order $n$ between directly *J*-coupled spins (with couplings above a user-specified threshold) and up to order $k$ between spatially proximate spins (with distances below the user-specified threshold). Generated by coupling graph analysis as described in the main text. The minimum basis set recommended for liquid state protein NMR simulations is IK-1(4,3) with the distance threshold of 4.0 Angstrom. |
| IK-2($n$) | For each spin, all of its correlations with directly *J*-coupled spins, and correlations up to order $n$ with spatially proximate spins (below the user-specified distance threshold). Generated by coupling graph analysis as described in the main text. Recommended for very accurate simulations on very large computer systems with the distance threshold of 5.0 Angstrom or greater. |

**Table 2:** CPU time and memory utilization statistics for ubiquitin $^1$H-$^1$H NOESY simulations shown in Figure 1 at different accuracy levels.

| Basis set for the reduced state space (see Table 1) | IK-1(2,2) | IK-1(3,2) | IK-1(4,2) | IK-1(4,3) |
|---|---|---|---|---|
| Reduced state space dimension | 29k | 56k | 210k | 849k |
| Number of non-zeroes in Hamiltonian superoperator | 43k | 223k | 1,420k | 2,500k |
| Number of non-zeroes in relaxation superoperator | 102k | 142k | 360k | 1,800k |
| Wall clock time (16 Sandy Bridge cores at 2.4 GHz) | 20 min | 58 min | 8 hours | 24 hours |

FIGURE 1

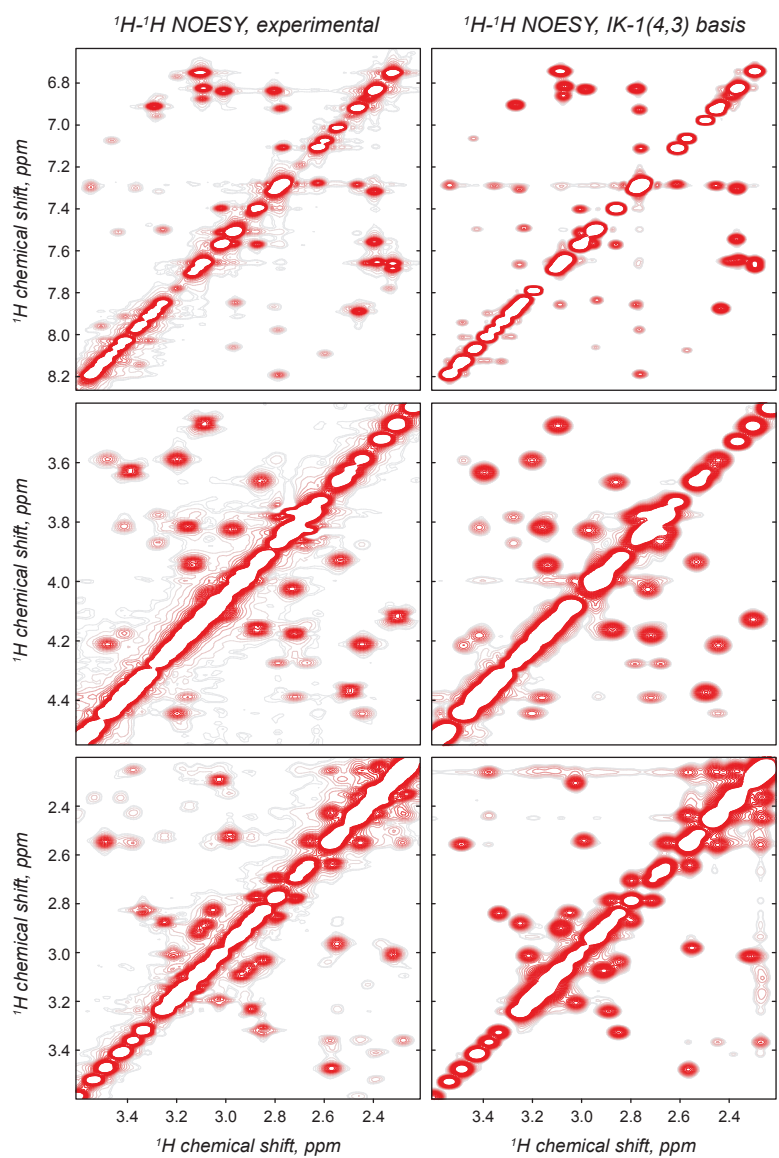

FIGURE 2

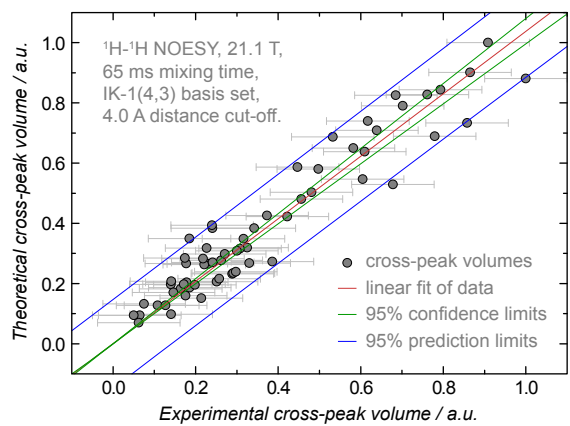

FIGURE 3

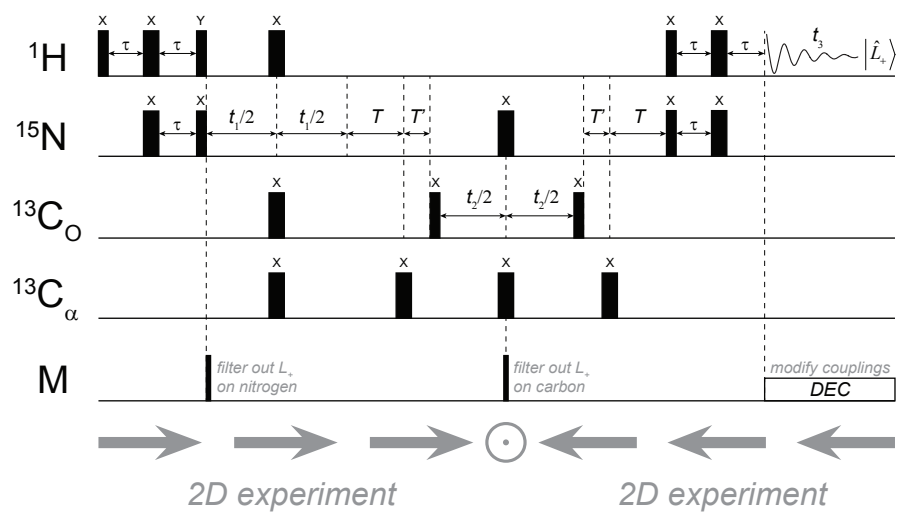

FIGURE 4

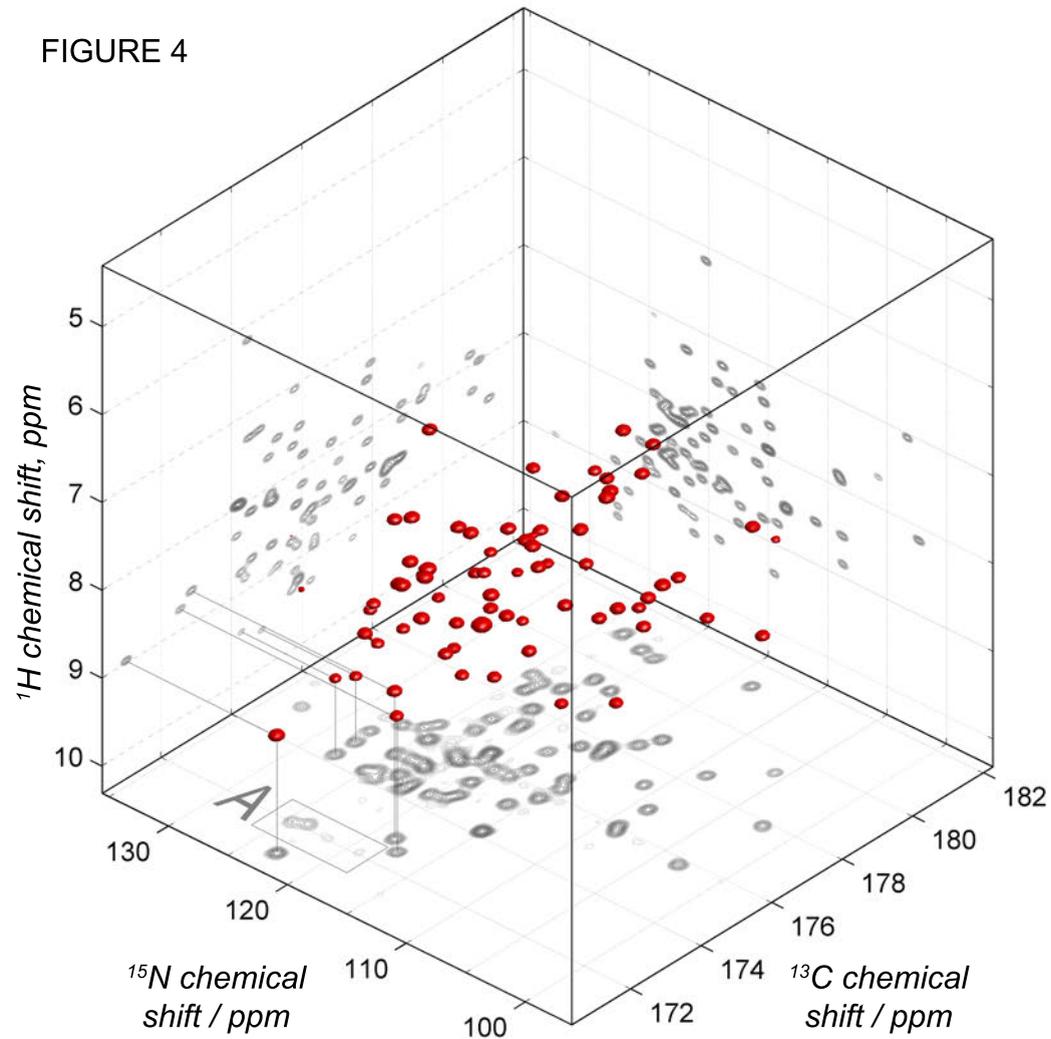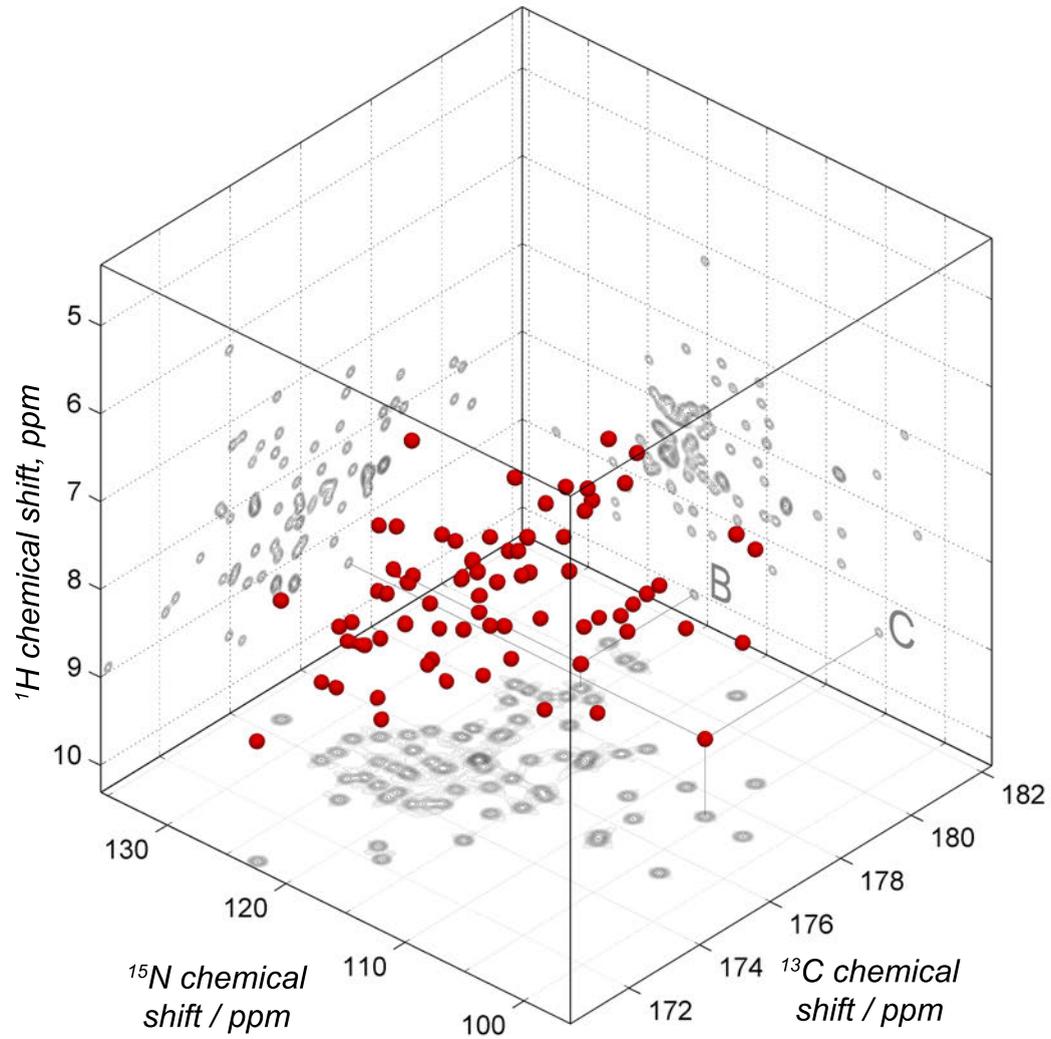

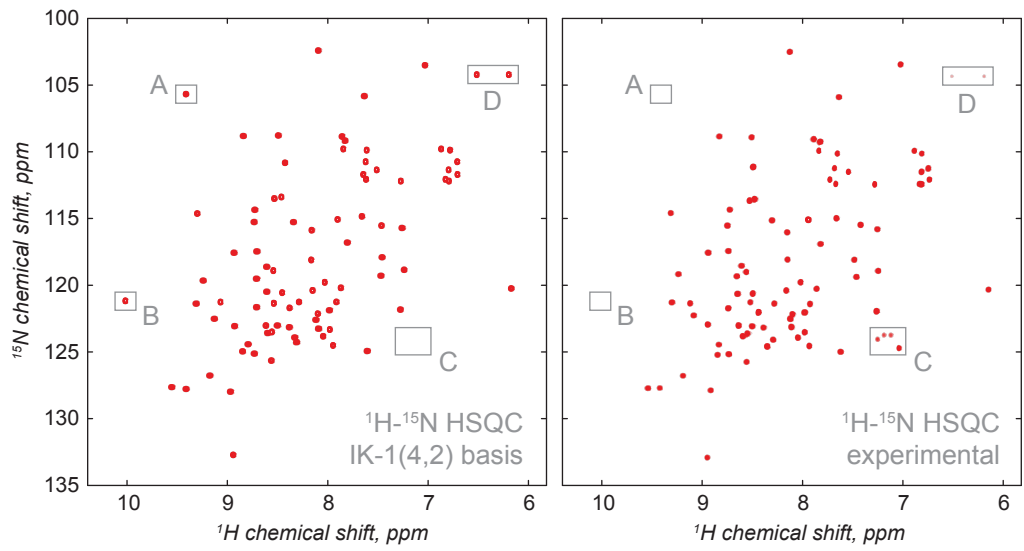